\documentclass{PoS}
\pdfoutput=1 
\usepackage{cancel}
\usepackage{slashed}
\usepackage{amsmath}
\usepackage{amssymb}
\usepackage{amsfonts}
\usepackage[english]{babel}

\newcommand{\alfive}{{\alpha_5}}
\newcommand{\CP}{{CP}}
\newcommand{\CPbar}{{\overline{CP}}}
\newcommand{\CPviol}{{\cancel{CP}}}

\newcommand{\mcL}{{\mathcal L}}
\newcommand{\mcO}{{\mathcal O}}
\newcommand{\mcE}{{\mathcal E}}
\newcommand{\Dslash}{{\slashed{D}}}

\title{Calculation of Nucleon Electric Dipole Moments 
  Induced by Quark Chromo-Electric Dipole Moments}

\ShortTitle{Neutron EDM induced by chromo-EDM}

\author{
  \speaker{Taku Izubuchi}${}^{a}$,
  Michael~Abramczyk${}^{b}$,
  Tom~Blum${}^{a,b}$,
  Hiroshi~Ohki${}^{a}$,
  Sergey~Syritsyn${}^{a,c,d}$
  \\
  \llap{${}^a$} RIKEN/BNL Research Center, Brookhaven National Laboratory, 
      Upton, NY 11973, USA\\
  \llap{${}^b$} Physics Department, University of Connecticut, 
      Storrs, CT 06269, USA\\
  \llap{${}^c$} Department of Physics and Astronomy, Stony Brook University, 
      Stony Brook, NY 11794, USA \\
  \llap{${}^d$} Thomas Jefferson National Accelerator Facility, 12000 Jefferson Avenue, 
      Newport News, VA 23606, USA
E-mail: \email{izubuchi@quark.phy.bnl.gov}}

\abstract{
We present initial results of computing nucleon electric dipole moment induced by quark
chromo-EDM, CP-violating quark-gluon coupling.
Using chirally-symmetric domain wall and M\"obius fermions with pion mass 
$m_\pi=172\text{ MeV}$, we calculate the connected part of the electric dipole 
form factor $F_3(Q^2)$.
In addition, we perform an exploratory study of the method to calculate EDM
using uniform background electric field on a lattice introduced without 
breaking the periodicity in the time direction.
}

\FullConference{34th annual International Symposium on Lattice Field Theory\\
		24-30 July 2016\\
		University of Southampton, UK}

\begin{document}

%%%%%%%%%%%%%%%%%%%%%%%%%%%%%%%%%%%%%%%%%%%%%%%%%%%%%%%%%%%%%%%%%%%%%%%%%%%%%%%
\section{Introduction}
The origin of nuclear matter, or the excess of nucleons over antinucleons 
in the early Universe, is one of the greatest puzzles
in physics known as the baryonic asymmetry of the Universe (BAU).
One of the required conditions for the BAU is violation of the $\CP$ symmetry.
Standard Model~(SM) $\CPviol$ from CKM matrix phases is not sufficient to explain BAU,
and signs of additional $\CPviol$ are actively sought in experiments.
The most promising ways to look for $\CPviol$ are measurements of
electric dipole moments (EDM) of atoms, nucleons and nuclei. 
Planned EDM experiments will improve their bounds by 2 orders of magnitude.
%Many extensions of SM admit $\CPviol$, and, given 
%that SM requires at least some extension to explain neutrino masses and
%possibly Dark Matter, searching for symmetry violations may yield insights into
%what these extensions may be.
Knowledge of nucleon structure and interactions is required to interpret these 
experiments in terms of quark and gluon fields and put constraints 
on many potential extensions of particle physics beyond the Standard Model (BSM) 
such as SUSY and GUT as sources of additional $\CPviol$.
Connecting the quark- and hadron-level effective $\CPviol$ 
interactions is a task for lattice QCD (an extensive
review of EDM phenomenology can be found in Ref.~\cite{Engel:2013lsa}).

Proton and neutron electric dipole moments may be induced by $\CPviol$ operators.
The only dim-4 operator is the QCD $\bar\theta$-angle.
The $\bar\theta$-induced nucleon EDMs have been calculated on a lattice previously 
from energy shifts in background electric 
field~\cite{Aoki:1989rx,Shintani:2006xr,Shintani:2008nt}
or extracting $\CPviol$ electric dipole form factor (EDFF) $F_3(Q^2)$~\cite{
Berruto:2005hg,Shintani:2005xg,
%Aoki:2008gv,
Shindler:2015aqa,
Shintani:2015vsx,Alexandrou:2015spa}.
%The latter method requires some form of extrapolation to $Q^2\to0$ and may be subject to
%systematic uncertainty.
%In all the calculations with the background electric field, the periodic boundary condition was
%broken due to the Wick rotation of the electric field.
%
Nucleon EDMs (nEDMs) induced by dim-6 operators have been studied using QCD sum rules and chiral
perturbation theory~\cite{Engel:2013lsa}.
%Quark EDM-induced nEDMs 
%are equivalent to the nucleon tensor ``charge'' and 
%have been recently
%computed on a lattice in partially-quenched framework~\cite{Bhattacharya:2015esa}.
Another important dim-6 operator is quark chromo-electric dipole moment (cEDM)
\begin{equation}
\label{eqn:qcedm}
\mcL_{cEDM} = i \sum_{\psi=u,d} \frac{\tilde\delta_\psi}{2}
    \bar\psi (T^a G^q_{\mu\nu}) \sigma^{\mu\nu}\gamma_5 \psi\,,\\
\end{equation}
and calculations of cEDM-induced nEDMs have started using Wilson
fermions~\cite{Bhattacharya:2016oqm}.

In this work, we present initial calculations of nEDM induced by quark chromo-EDM using
chirally-symmetric Domain wall fermions.
We compute connected parts of the EDFF $F_3$ from 4-point correlation functions of nucleon with
quark current and cEDM quark-bilinear operator~(\ref{eqn:qcedm}).
In addition, we test the method to compute EDM from energy shift in background electric field,
which is introduced without violating boundary conditions following Ref.~\cite{Detmold:2009dx},
on a small $16^3\times32$ lattice and find ballpark agreement between the two methods\footnote{
  Preliminary results presented in the conference poster did not show the form factor dependence 
  on momentum transfer and energy shift method. 
  We omit gradient flow study from this manuscript because of space considerations.
}.
%Finally, we study effects of gradient flow applied to cEDM operator on the signal-to-noise of
%cEDM-nucleon correlation functions.

%%%%%%%%%%%%%%%%%%%%%%%%%%%%%%%%%%%%%%%%%%%%%%%%%%%%%%%%%%%%%%%%%%%%%%%%%%%%%%%
\section{Lattice details}
In our preliminary calculation, we employ two lattice ensembles with $N_f=2+1$ dynamical domain
wall fermions listed in Tab.~\ref{tab:ens}.
The first is a DSDR ensemble~\cite{Arthur:2012yc} with light pion mass $m_\pi\approx172\text{
MeV}$, on which we calculate the electric dipole form factor (EDFF) $F_3(Q^2)$ 
from vector current matrix elements of the nucleon in $\CPviol$ vacuum. 
The forward limit of the EDFF gives the EDM, $d_N=F_3(0)/(2m_N)$.
%, mainly as a feasibility study of the statistical precision for later calculation.
%We also study the effects of gradient flow on the chromo-EDM operator on these lattices.
The second is a Iwasaki+DWF ensemble with heavier pion mass 
$m_\pi\approx420\text{ MeV}$~\cite{Blum:2011pu}, on which we calculate 
the nucleon EDFF from vector current matrix elements and the nucleon EDM 
from nucleon energy shifts in background electric field.
%Calculations on the second ensemble are much cheaper and we 
%can afford substantially more statistics to perform comparison of
In large-statistics calculations on the second ensemble, we aim to compare the EDFF and EDM 
computed with the two different methods.
%from the form factor and the energy shift.
%, although smaller volume ($16^3$ vs. $32^3$) somewhat reduces the statistical precision.

\begin{table}[ht!]
\caption{Lattice ensembles on which the simulations were performed. 
  The statistics are shown for ``sloppy'' (low-precision) samples.
  \label{tab:ens}}
\begin{tabular}{ll|cccc|rrrr}
\hline\hline
 & $L_x\times L_t\times L_5$ 
  & $a\text{ [fm]}$ & $am_l$ & $am_s$ &  $m_\pi\text{ [MeV]}$ 
  & conf & stat & $N_{ev}$ & $N_{CG}$ \\
\hline
DSDR & $32^3\times64\times32$ & 
  0.1432(7) & 0.001 & 0.045 & 172.3(9) & 28 & 896 & 1400 & 125 \\
Iwasaki & $16^3\times32\times16$ &  
  0.114(2) & 0.01 & 0.032 & 422(7) & 500 & 16500 & 150 & 100 \\
\hline\hline
\end{tabular}
\end{table}

\begin{figure}[ht!]
\centering
\includegraphics[width=.4\textwidth,clip]{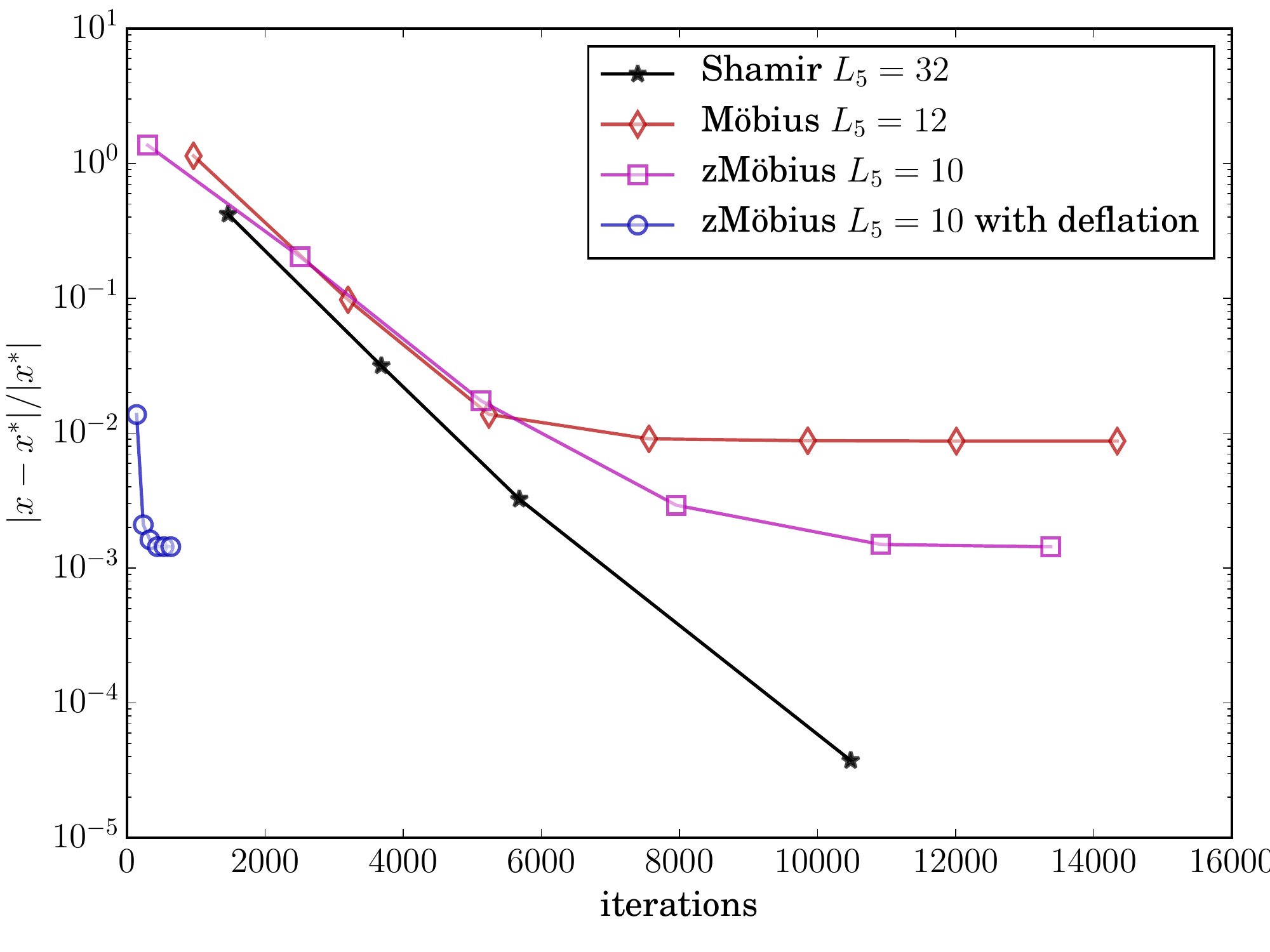}~
\hspace{.05\textwidth}~
\includegraphics[width=.4\textwidth,clip]{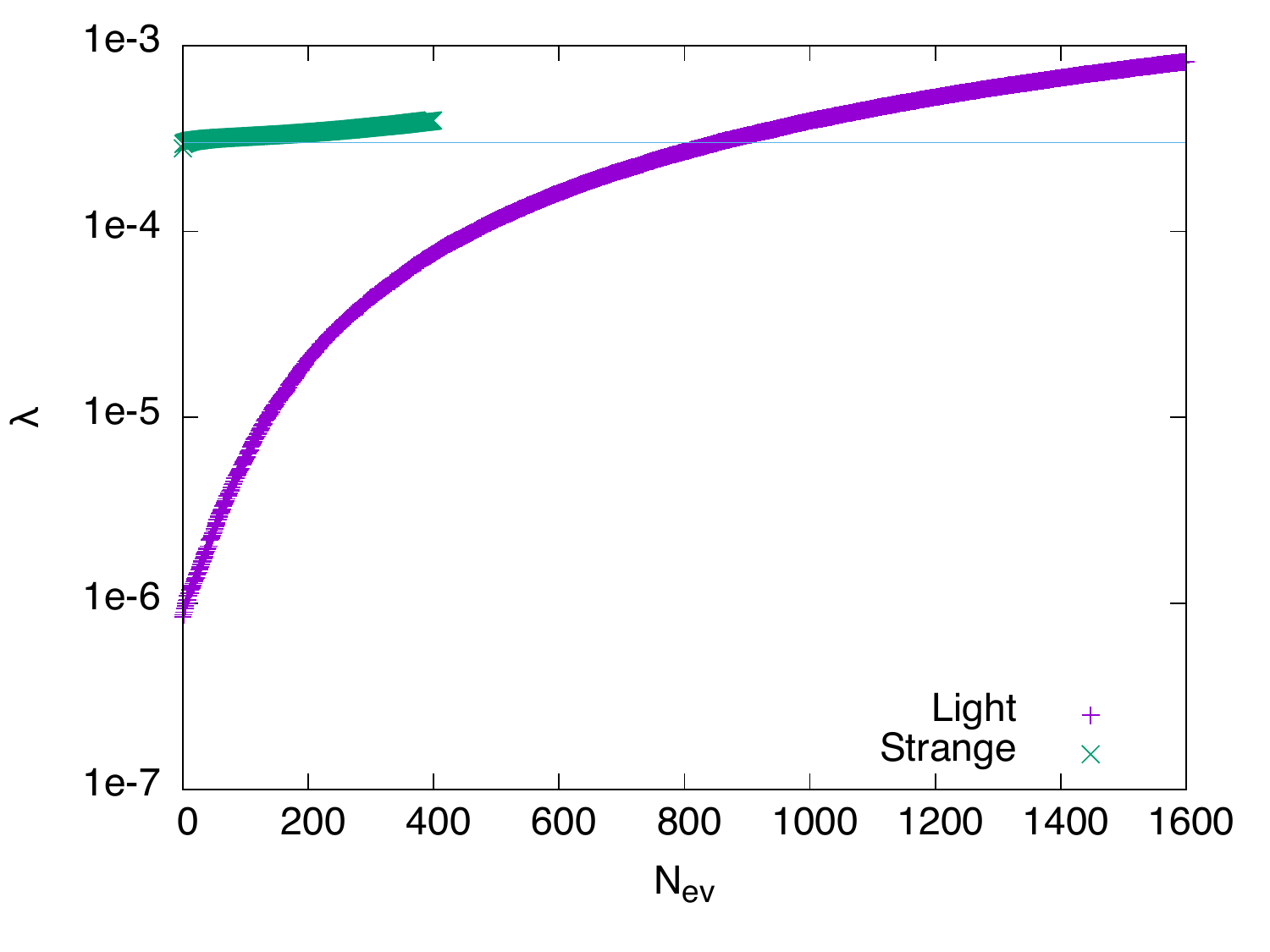}
\caption{ (Left) Comparison of 4d solution vectors (quark propagators) between approximate
  5D Dirac operators ($L_5=10$ complex ``zMobius'' and $L_5=12$ regular Mobius)
  and the exact Shamir quark propagator with $L_5=32$ vs the iteration count.
  (Right) Low-mode eigenvalues of zMobius $L_5=10$ computed with light and strange 
  quark masses. \label{fig:mobius-study}}
\end{figure}

%In order to accelerate the calculations, 
We use \emph{all-mode-averaging}~\cite{Shintani:2014vja} framework to optimize sampling, 
in which we  approximate quark propagators with truncated-CG solutions 
to a M\"obius operator~\cite{Brower:2005qw}.
We use the M\"obius operator with short 5th dimension $L_{5s}$ and complex $s$-dependent
coefficients $b_s + c_s = \omega_s^{-1}$ (later referred to as ``zMobius'') that approximates 
the same 4d effective operator as the Shamir operator with the full $L_{5f}=32$ (DSDR) 
or $L_{5f}=16$ (Iwasaki).
The approximation is based on the domain wall-overlap equivalence
\begin{gather}
[\Dslash^\text{DWF}]_{4d} = \frac{1+m_q}2 - \frac{1-m_q}2 \gamma_5
\epsilon_{L_5}(H_T)\,,
\quad
H_T = \gamma_5\frac{\Dslash_W}{2+\Dslash_W}\,,\\
\epsilon^\text{M\"obius}_{L_{5s}}(x) 
  = \frac{\prod_s^{L_{5s}}(1+\omega_s^{-1}x) - \prod_s^{L_{5s}}(1-\omega_s^{-1}x)}
         {\prod_s^{L_{5s}}(1+\omega_s^{-1}x) + \prod_s^{L_{5s}}(1-\omega_s^{-1}x)}
  \approx \epsilon^\text{Shamir}_{L_{5f}}(x) %= \frac{(1+\omega_s^{-1}x)}
  \,.
\end{gather}
where the coefficients $\omega_s$ are chosen so that the function 
$\epsilon^\text{M\"obius}_{L_{5s}}(x)$ is the \emph{minmax} approximation to the 
$\epsilon^\text{Shamir}_{L_{5f}}(x)$.
We find that $L_{5s}=10$ is enough for efficient 4d operator approximation.
Shortened 5th dimension reduces the CPU and memory requirements, which is especially significant
for the lighter pion mass ensemble $L_{5f}=32$ is reduced to $L_{5s}=10$ saving $70\%$ of
the cost.
We deflate the low-lying eigenmodes of the the internal even-odd preconditioned operator,
to make truncated-CG AMA efficient. 
The numbers of deflation eigenvectors $N_{ev}$ and truncated CG iterations $N_{CG}$ are given in
Tab.~\ref{tab:ens}.
Figure~\ref{fig:mobius-study}~(left) shows the deviation of truncated solutions to 
the zM\"obius operator from the exact Shamir solution, demonstrating the effect of deflating 
the eigenmodes between the light and the strange quark masses 
(Fig.~\ref{fig:mobius-study}, right).
We compute 32 sloppy samples per configuration. 
To correct any potential bias due to the approximate $\Dslash$ operator and truncated CG, 
we compute one exact sample with the Shamir operator per configuration.
The latter is computed iteratively by refining the solution of the ``zMobius'' 
to approach the solution of the Shamir operator, again taking advantage of the short $L_{5s}$
and deflation.

%%%%%%%%%%%%%%%%%%%%%%%%%%%%%%%%%%%%%%%%%%%%%%%%%%%%%%%%%%%%%%%%%%%%%%%%%%%%%%%
\section{Electric Dipole Form factor $F_3$}

\begin{figure}[t]
\centering
\includegraphics[width=.8\textwidth]{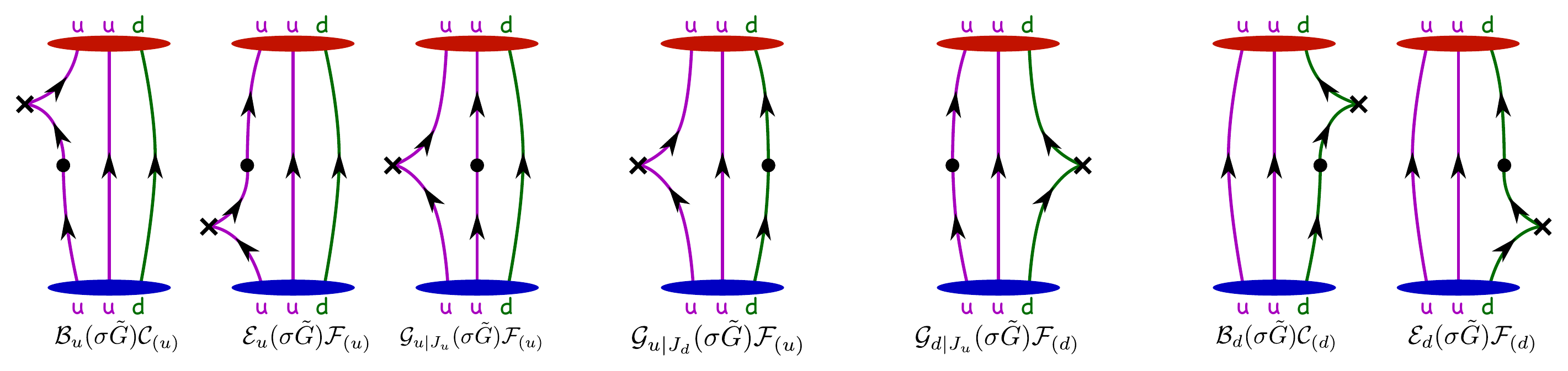}
\caption{Quark-connected contractions of nucleon, quark current, and cEDM operators. 
  \label{fig:cedm_contract}}
\end{figure}

The EDFF $F_3$ is a parity-odd quantity induced by $\CPviol$ interactions.
To compute the effect of chromo-EDM~(\ref{eqn:qcedm}), we Taylor-expand QCD$+\CPviol$ 
vacuum averages as
$S\to S + i\sum_x \tilde\delta_\psi \mcO^{\CPbar}_{\psi,x}$
\begin{gather}
\label{eqn:corr_cpviol}
\langle N\,[\bar q \gamma_\mu q]\, \bar N \rangle_{\CPviol}
  = C_{NJ\bar N} - i\tilde\delta_\psi \,\delta_\psi^\CPbar C_{NJ\bar N}
    + O(\tilde\delta_\psi^2)\,,
\\
\nonumber
\text{with }\quad
C_{NJ\bar N} = \langle N\,[\bar q \gamma_\mu q]\, \bar N\rangle\,,
\quad \delta_\psi^\CPbar C_{NJ\bar N} 
= \langle N\,[\bar q \gamma_\mu q]\, \bar{N} \cdot \sum_x\mcO^{\CPbar}_{\psi,x} \rangle\,,
\quad
\mcO_\psi^{\CPbar}  = \frac12\bar\psi (T^aG^a_{\mu\nu})\sigma^{\mu\nu}\gamma_5\psi\,,
%\langle\mcO\ldots\rangle_{\CPviol} = \langle\mcO\ldots\rangle 
%  - i\tilde d\langle\mcO^{\CPbar}\mcO\ldots\rangle + O(\tilde d^2)
\end{gather}
where $\langle\cdot\rangle$ in the second line stand for vacuum averages computed 
with $\CP$-even action $S$.
In this work, we calculate only fully connected contributions to the correlation 
functions $C_{N J \bar N}$ and $\delta^{\CPbar} C_{N J\bar N}$.
To compute the latter, the quark-bilinear cEDM density~(\ref{eqn:qcedm}) is inserted 
once in every $\psi$-quark line of $C_{N J\bar N}$ diagrams, generating the four-point functions 
shown in Fig.\ref{fig:cedm_contract}.
We evaluate these 4-point contractions using sequential propagators.
In addition to the usual one forward and two backward (sink-sequential) propagators, we compute 
one cEDM-sequential and four doubly-sequential (\{cEDM, sink\}-sequential) propagators.
This is the minimum number inversions required to compute cEDM-induced nEDM for all combinations 
of two degenerate flavors $u$ and $d$, with one source-sink separation $t_\text{sep}$ 
and one sink momentum $\vec p^\prime$. 
Compared to the method used in Ref.~\cite{Bhattacharya:2016oqm}, in effect we compute the first 
derivative in the $\CP$-odd coupling $\epsilon$ at $\epsilon=0$, obviating any higher-order 
dependence on $\epsilon$.
We use $\vec p^\prime=0$ and $t_\text{sep}=8a=1.14\text{ fm}$ for the $32^3\times64$ and
$t_\text{sep}=\{8,10\}a=0.91,1.15\text{ fm}$ for the $16^3\times32$ ensemble.
The $\CP$-even part of the nucleon-current correlator~(\ref{eqn:corr_cpviol})
requires only contractions but no additional inversions.
Both $C_{NJ\bar N}$ and $\delta^{\CPbar}C_{NJ\bar N}$ are computed with
the polarization projector 
$T_\text{pol}=\frac{1+\gamma_4}2\Sigma_3=\frac{1+\gamma_4}2(-i\gamma_1\gamma_2)$.
This projector is sufficient to extract form factors $F_{1,2}$ and $F_3$ 
from $\CPviol$ nucleon matrix elements of the vector current\footnote{
  All conventions for form factors and momenta are Euclidean}
\begin{equation}
\langle N_{p^\prime} | \bar{q}\gamma^\mu q | N_p\rangle
  = \bar{u}_{p^\prime} \big[ F_1(Q^2) \gamma^\mu
      + F_2(Q^2) \frac{\sigma^{\mu\nu} q_\nu}{2m_N}
      + F_3(Q^2) \frac{i\gamma_5\sigma^{\mu\nu} q_\nu}{2m_N}
      \big] u_p\,.
\end{equation}
%The on-shell spinors $u_p$, $\bar{u}_{p^\prime}$ satisfy the usual Dirac equation.

\begin{figure}[t]
\centering
\vspace{-.2cm}
\includegraphics[width=.40\textwidth]{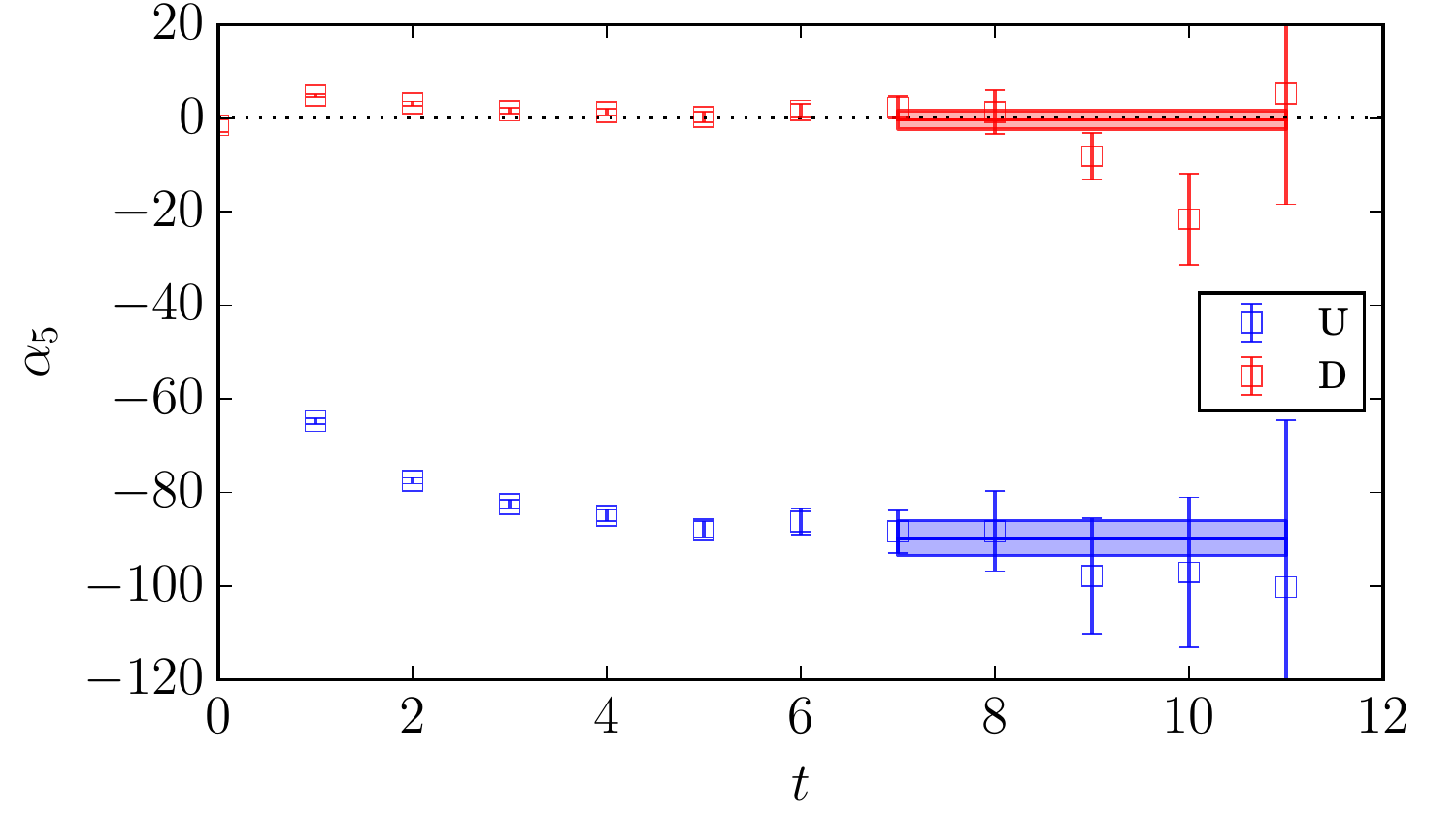}~
\hspace{.05\textwidth}~
\includegraphics[width=.40\textwidth]{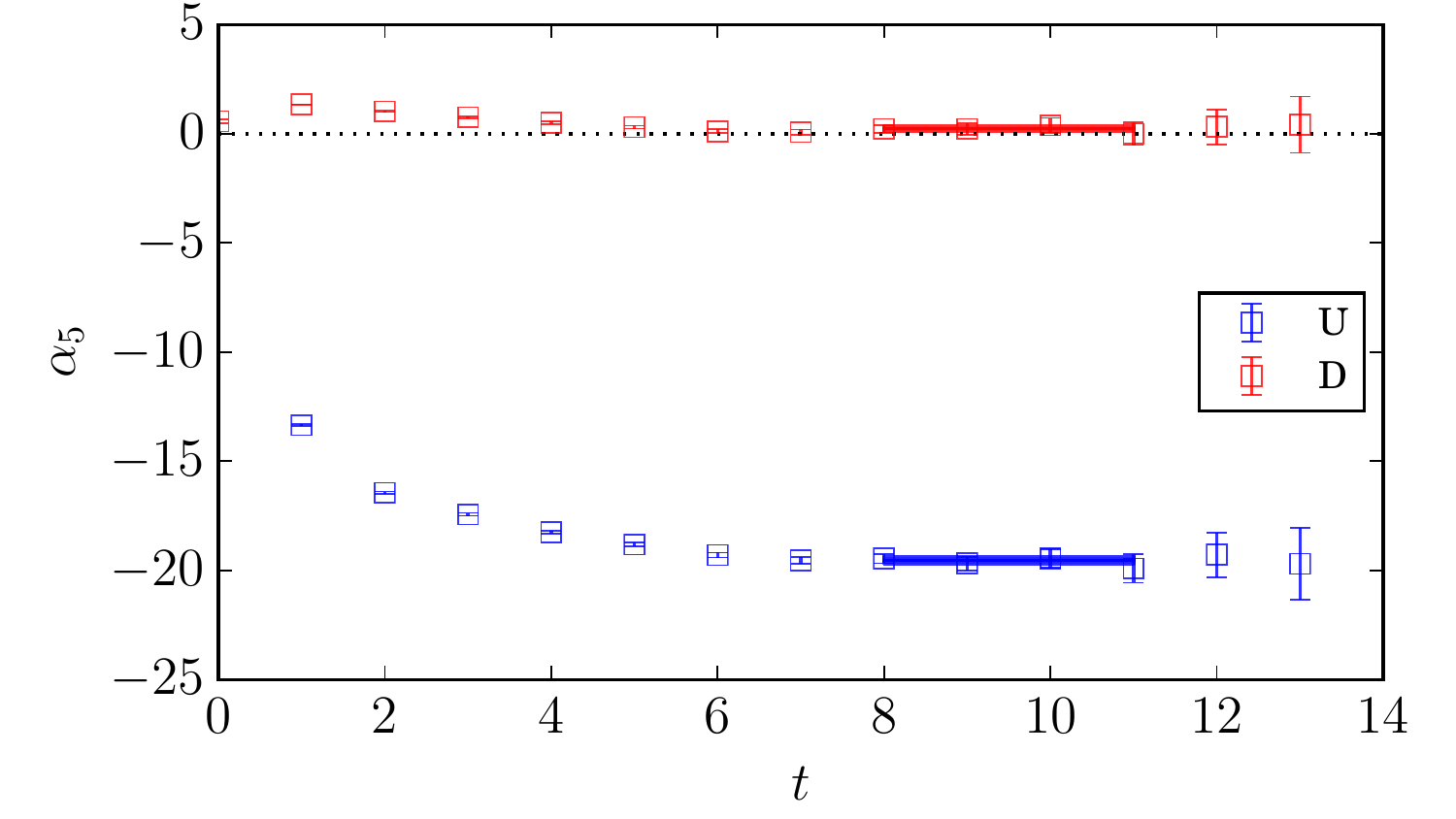}\\
\vspace{-.4cm}
\caption{The parity-mixing phase of the proton propagator induced by $u$- and $d$-cEDM, for the
  $m_\pi=172$ (left) and $422\text{ MeV}$ (right). 
  The mixing is noticeably larger when the unpaired quark is affected by cEDM.
  \label{fig:alfive}}
\end{figure}

\begin{figure}[t]
\centering
\vspace{-.2cm}
\includegraphics[width=.40\textwidth]{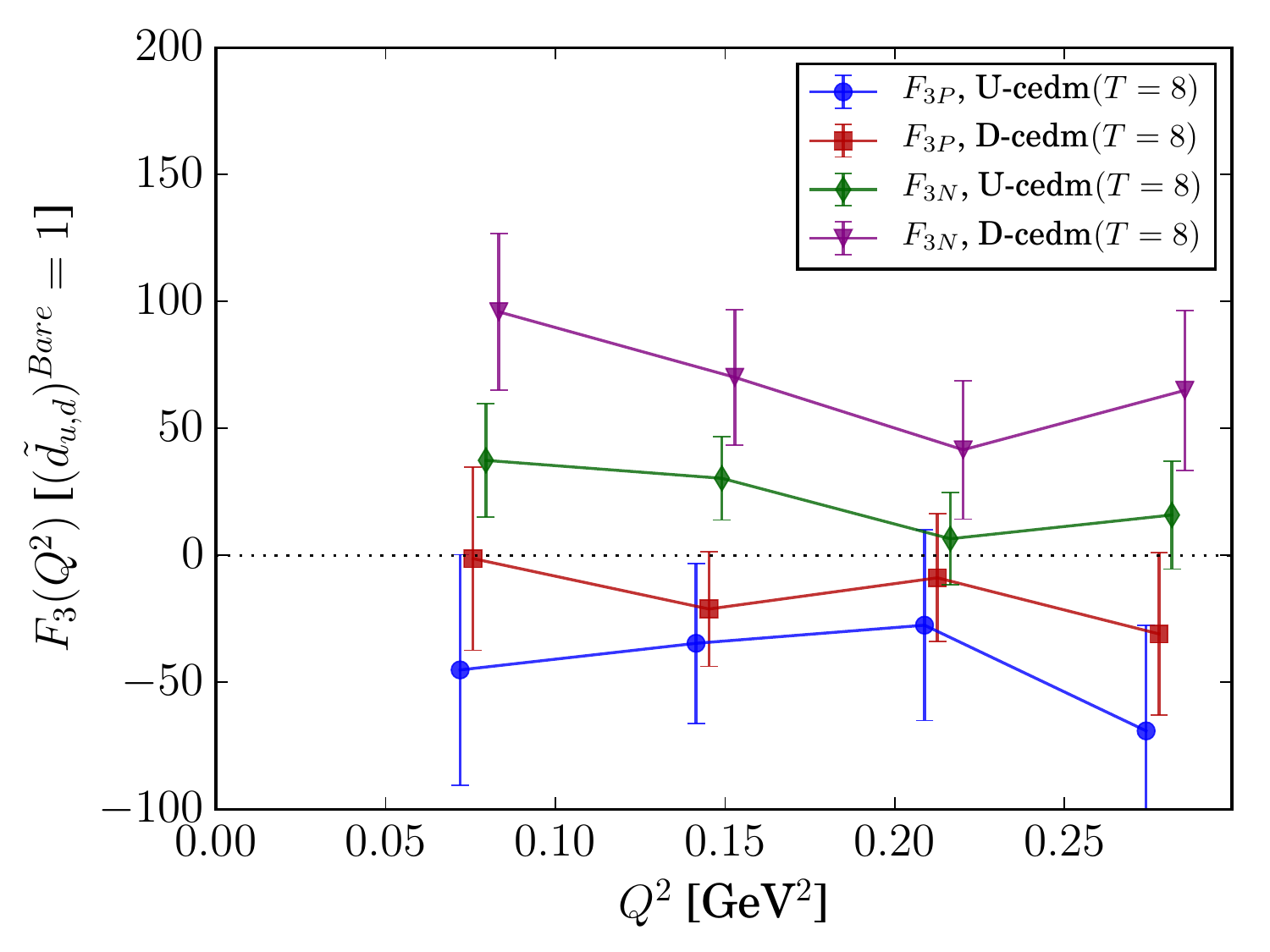}~
\hspace{.05\textwidth}~
\includegraphics[width=.40\textwidth]{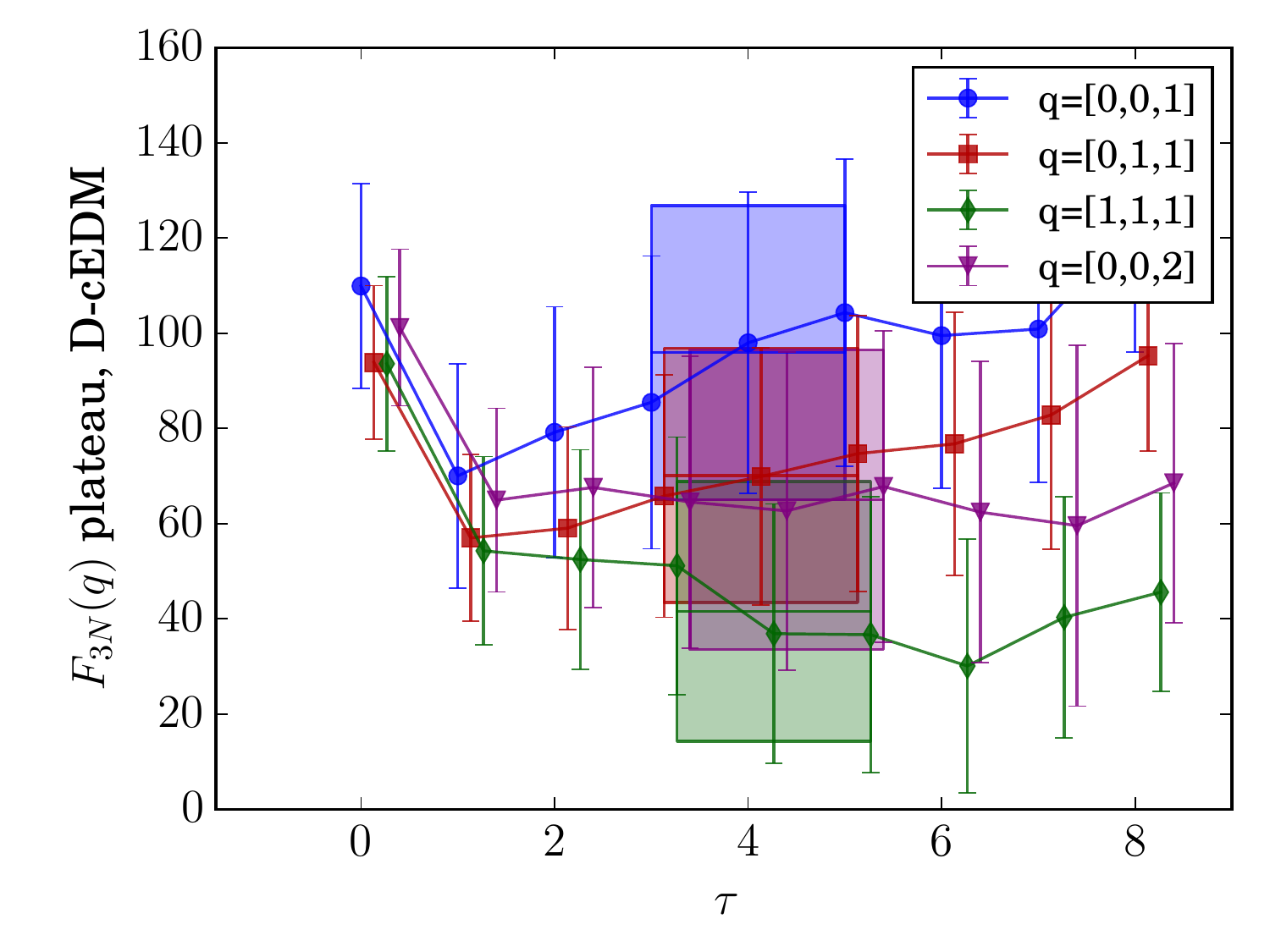}\\
\vspace{-.4cm}
\caption{Form factors computed on $m_\pi=172$ ensemble: proton and neutron EDFF 
  $F_3(Q^2)$ induced by $u$-  and $d$-cEDM (left)
  and form factor plateaus for neutron $F_3$ induced by $d$-quark cEDM (right).
  Source-sink separation is $t_\text{sep}=8a=1.15\text{ fm}$.
  \label{fig:ff_32c64}}
\end{figure}

We employ the usual $C_{3pt}/C_{2pt}$ ratios to cancel overlaps and exponential factors 
and we extract matrix elements of the ground state as the central plateau average (see
Fig.~\ref{fig:ff_32c64}, right).
Excited state contaminations are neglected in this preliminary analysis.
The P-even form factors $F_{1,2}$ are extracted from overdetermined fit (e.g.,
Ref.~\cite{Syritsyn:2009mx}) of $C_{NJ^\mu\bar N}$ combining all orientations 
of the current and momenta.
The EDFF $F_3$ is extracted from the timelike-component of 
$\delta^{\CPbar}C_{NJ^4\bar N}$.
$\CPviol$ interaction leads to parity mixing of the nucleon spinors
\begin{equation}
\tilde u_p = e^{i\alfive\gamma_5} u_p, 
\quad
\langle N_\delta(t,\vec p) \bar N_{\delta^\prime}(0)\rangle 
  \stackrel{t\to\infty}= \frac{[-i\slashed{p} 
        + m_N e^{2i\alfive\gamma_5}]_{\delta\delta^\prime}}{2E_N} e^{-E_N t}
\end{equation}
that complicates analysis of $F_3$, as $\delta^{\CPbar}C_{NJ^4\bar N}$ receives contributions 
from $F_{1,2}$ due to ``polarization mixing'' 
$\sim\alfive\{\gamma_5,T_\text{pol}\}$~\cite{Abramczyk:2017oxr}.
To subtract them, we determine the cEDM flavor-dependent mixing angle from the $\CP$-odd 
two-point function $\delta^{\CPbar}C_{N\bar N}$ as 
(see Fig.~\ref{fig:alfive}) 
\begin{equation}
\alpha_5(t) 
  = - { \mathrm{Re}\mathrm{Tr}\big[T^+ \gamma_5 \,  \delta^\CPbar C_{N\bar N}(t)\big]} 
    /  { \mathrm{Re}\mathrm{Tr}\big[T^+ \, C_{N\bar N}(t)\big] }\,,
\end{equation}
where $T^+=(1+\gamma_4)/2$ is the positive-parity projector.
Remarkably, connected cEDM leads to parity rotation only when $\CPviol$ interaction affects the
``spectator'' (non-diquark) flavor in the nucleon.
We use the value $\alfive(t_\text{sep})$ and $\tau$-dependent plateau values of $F_{1,2}$ 
for the subtraction. 
The subtracted formula for $F_3$ is~\cite{Abramczyk:2017oxr}
\begin{equation}
F_3^\psi = \frac{2m_N}{E_N + m_N} \Big[
  \frac{2E_N(E_N+m_N)}{q_3}\mathrm{Tr}\big[T^+ \Sigma_z 
      \delta^{\CPbar}_\psi R_{N J^4 \bar{N}}\big]
  -\alfive G_E \Big]\,,
\end{equation}
where spin $\Sigma_z=-i\gamma_1\gamma_2$ and 
$\delta^{\CPbar} R_{N J^4 \bar{N}}$ is the $\delta^{\CPbar}C_{3pt}/C_{2pt}$ ratio canceling 
nucleon field normalization and time exponentials.
The $F_3$ results for the $m_\pi=172\text{ MeV}$ ensemble are shown in 
Fig.~\ref{fig:ff_32c64}\footnote{
  On one out of 30 analyzed configurations, we have encountered an outlier sample that had
  values for the 2- and 3-point functions strongly deviating from the rest of the ensemble. 
  We have omitted the entire gauge configuraion as well as one adjacent to it from the 
  preliminary analysis, and we are investigating what caused this outlier.
}

%%%%%%%%%%%%%%%%%%%%%%%%%%%%%%%%%%%%%%%%%%%%%%%%%%%%%%%%%%%%%%%%%%%%%%%%%%%%%%%
\section{Energy shift in background electric field}
The background field method provides opportunity to compute EDMs directly, whereas 
$Q^2\to0$ extrapolation of $F_3(Q^2)$ may lead to systematic uncertainties.
We have done an exploratory calculation of cEDM-induced nEDM on the $16^3\times32$
$m_\pi=422\text{ MeV}$ ensemble in uniform electric field to compare these two methods.
We use the Wick-rotated electric field, which, if quantized as 
$Q_q \mcE_k=n_k\frac{2\pi}{L_k L_t}$, does not break (a)periodic boundary
conditions~\cite{Detmold:2009dx}.
The Wick-rotated electric field cannot create pion pairs out of vacuum, but 
it can ``accelerate'' charged particles and complicate calculation of their rest mass.
For this reason, we apply this analysis only to the neutron.

We compute the energy shift from the $\CP$-odd correction $\delta^\CPbar C_{N\bar N,\mcE_z}$ 
to the neutron propagator in presence of the uniform electric field $\mcE_z$
\begin{equation}
U_{\mu}\to U_{\mu} e^{iQA_{\mu}}\,,
\quad
A_{\hat t}(z,t) = z\, \mcE_z \,\, \big(\forall z=0\ldots (L_z-1)\big)\,,
\quad
A_{\hat z}(z,t) = -t L_z\,\mcE \,\,\big(\forall t=L_t-1)\,.
\end{equation}
$\delta^\CPbar C_{N\bar N,\mcE_z}$ is computed similarly to Fig.~\ref{fig:cedm_contract}
but without the quark current.
$\CP$-odd interaction leads to spin-dependent energy shift
$\delta E = - d_N \, \hat S\cdot i\vec\mcE$, from which we extract $d_N$ as 
\begin{equation}
\label{eqn:dneff}
2 m_N d_N^\text{eff}(t) = -\frac{2m_N}{\mcE_z} \big[R_z(t+1) - R(t)\big]\,,
\quad R_z(t) = \frac{\mathrm{Tr}\big[ T^+ \,\Sigma_z\,\delta^\CPbar C_{N\bar N,\mcE_z}(t)\big]}
                    {\mathrm{Tr}\big[ T^+ \, C_{N\bar N,\mcE_z}(t)\big]}\,.
\end{equation}
We have computed $d_N^\text{eff}(t)$ for two values of the electric field ($1\times$ and $2\times$
$\mcE_\text{min}$), and averaged the result over $\mcE_z$ and $S_z$ directions.
The statistics are the same as in the form factor calculation.

\begin{figure}[ht!]
\centering
\vspace{-.2cm}
\includegraphics[width=.45\textwidth]{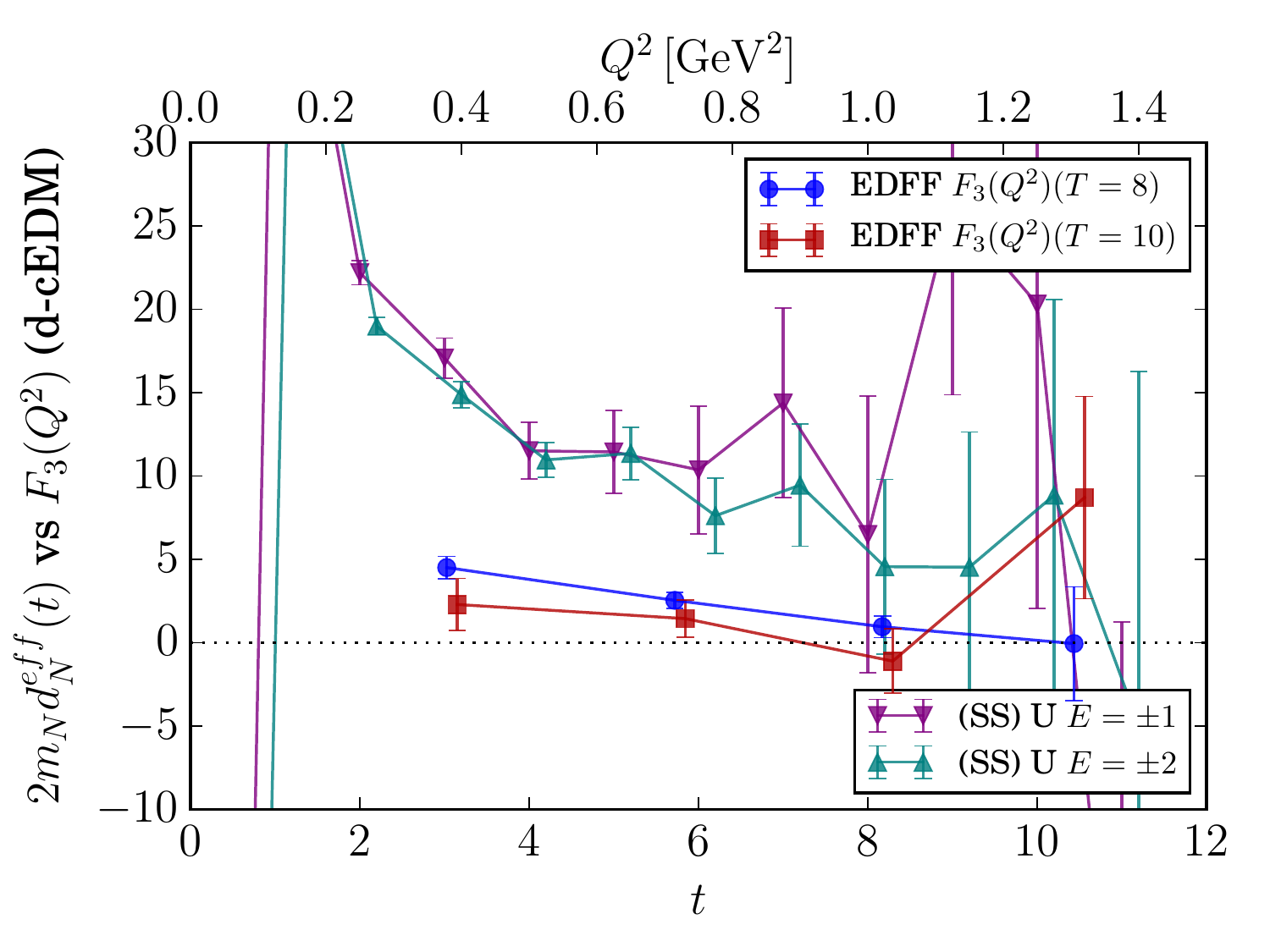}~
\hspace{.05\textwidth}~
\includegraphics[width=.45\textwidth]{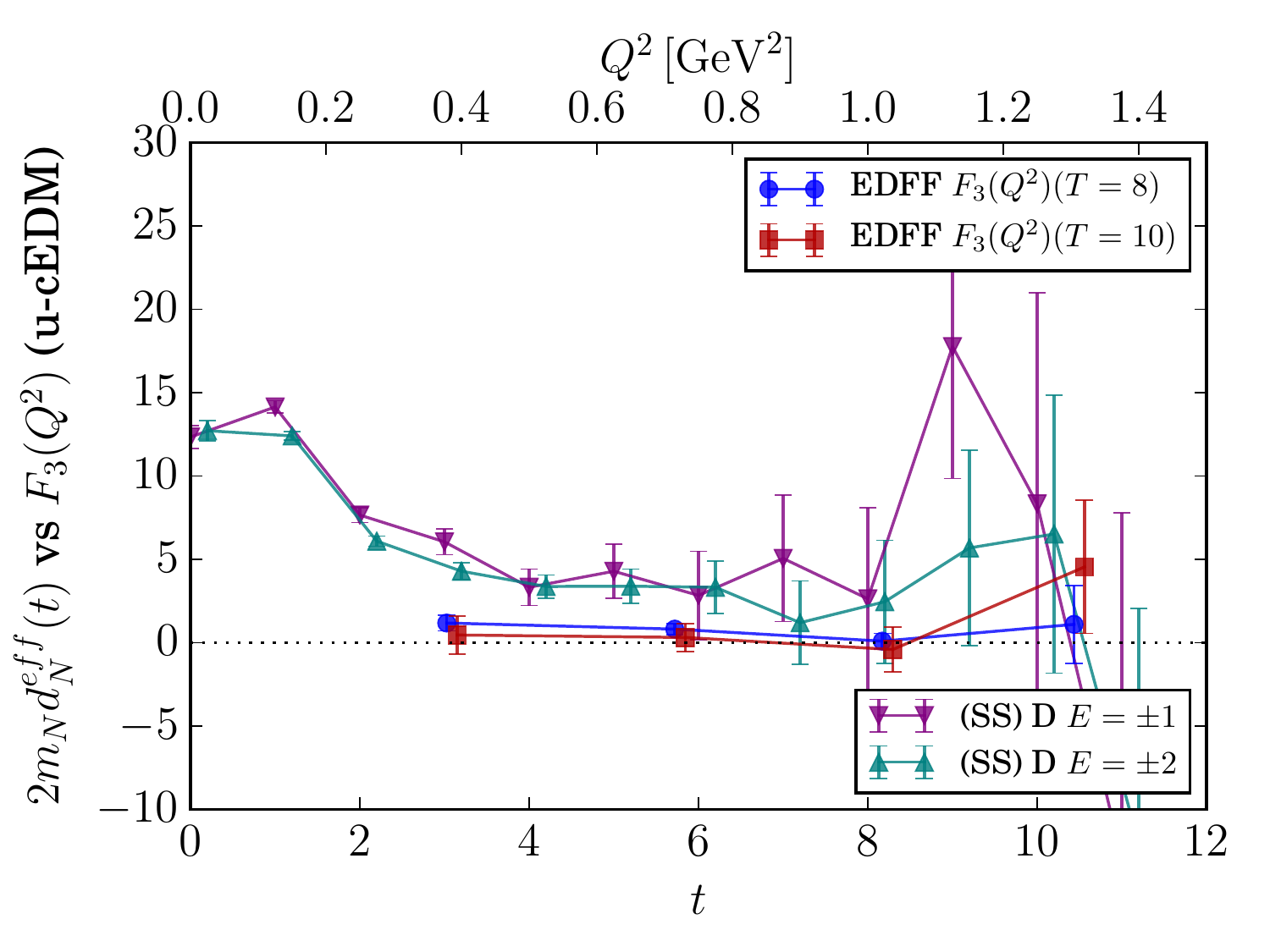}\\
\vspace{-.4cm}
\caption{Comparison of neutron EDFF $F_3(Q^2)$ vs. $Q^2$ (top scale) 
  effective EDM $2 m_N d_N^\text{eff}(t)$~(\ref{eqn:dneff}) vs. $t$ (bottom scale) computed from 
  energy shift on the $422\text{ MeV}$ ensemble, for $d$-cEDM (left) and $u$-cEDM(right).
  \label{fig:dneff_vs_f3} }
\end{figure}

In Figure~\ref{fig:dneff_vs_f3} we compare $F_3(Q^2)$ and $2m_N d_N^\text{eff}(t)$, where 
the EDM ``units'' are adjusted for a direct comparison.
%analysis for to check if they indeed agree: EDFF $F_3(Q^2)$ must be extrapolated to $Q^2\to0$
%and $d_N^\text{eff}(t)$ must be extrapolated to $t\to\infty$.
The EDM value from the energy shift appears to reach plateau for $t=4\ldots6$. 
However, the comparison EDFF values computed with $t_\text{sep}=8a$ and $10a$ shows that
$F_3$ requires detailed analysis of excited state contributions. 
For the $d$-cEDM, there appears to be $\times2\ldots\times2.5$ difference between
$d_N^\text{eff}$ and $F_3(Q^2_\text{min})$.
Considering the trend of $F_3(Q^2)$ with $Q^2\to0$, the results of form factor and energy shift 
calculations appear to be compatible.
We note that the minimal electric field is 
$\mcE_\text{min}=0.097\text{ GeV}^2=490\text{ MeV/fm}$ on the small $16^3\times32$ lattice.
Such strong electric field may distort the nucleon and introduce systematic shift to
$d_N^\text{eff}$, complicating the comparison.
Nevertheless, we find the data suggest encouraging agreement between the two methods.

\section{Summary}
We have obtained preliminary results for the nucleon electric dipole form factors induced 
by light quark chromo-EDM interaction using chirally-symmetric domain wall fermions.
Our initial calculation does not take into account renormalization and mixing of chromo-EDM 
operators with other $\CPviol$ operators, and neglects effects of excited states. 
%The electric dipole form factor shows significant excited states effect, which will need
%further study.
In addition, we have performed an exploratory study of electric dipole moments 
using the background electric field method and have compared them to the form factor results.
While the statistical and systematic precision is insufficient to draw definite conclusions,
these results appear to be comparable.
Using methods developed in this work, the studies will be extended to the physical point 
and will include analysis of excited states, as well as renormalization and mixing of 
$\CPviol$ operators.

\section*{ACKNOWLEDGEMENTS}
We are grateful to the RBC/UKQCD collaboration for the gauge configurations
and to the RIKEN ACCC facility for the computing resources provided for this work.
T.B. is supported by US DOE grant DE-FG02-92ER40716.
T.I. is supported in part by the Japanese Ministry of Education Grant-in-Aid, No.
26400261
The work of H.O. is supported by the RIKEN Special Postdoctoral Researcher program.
S.S. thanks RBRC for support under its joint tenure track fellowship with
Stony Brook University
S.S. also thanks KITP UCSB for hospitality at the ``Nuclear16'' workshop
and support by the National Science Foundation under Grant No. NSF PHY11-25915.

\bibliographystyle{aip}
\bibliography{cedm-lat16}

%\begin{thebibliography}{99}
%\bibitem{...}
%\end{thebibliography}

\end{document}